\documentclass[preprint,showpacs,showkeys,aps,12pt]{revtex4}
\usepackage{amsmath}
\usepackage{amssymb}
\usepackage{epsfig}
\usepackage{slashed}
\usepackage{graphicx}
\usepackage{multirow,color}
\graphicspath{{fig/}}
\begin{document}
\title{ Double Elementary Goldstone Higgs Boson Production in Future Linear Colliders}
\author{Yu-Chen Guo}\email{lgguoyuchen@126.com}
\author{Chong-Xing Yue}\email{cxyue@lnnu.edu.cn}
\author{Zhi-Cheng Liu}

\affiliation{
Department of Physics, Liaoning Normal University, Dalian 116029,  China
\vspace*{1.5cm}}

\begin{abstract}

The Elementary Goldstone Higgs (EGH) model is a perturbative extension of the standard model (SM), which identifies the EGH boson as the observed Higgs boson. In this paper, we study pair production of the EGH boson in future linear electron positron colliders. The cross sections in the TeV region can be changed to about $-27\%$, $163\%$ and $-34\%$ for the $e^+e^-\rightarrow Zhh$, $e^+e^-\rightarrow \nu\bar{\nu}hh$, and $e^+e^-\rightarrow t\bar{t}hh$ processes with respect to the SM predictions, respectively.  According to the expected measurement precisions, such correction effects might be observed in future linear colliders. In addition, we compare the cross sections of double SM-like Higgs boson production with the predictions in other new physics models.

\end{abstract}



\maketitle
\section{Introduction}

Despite the discovery of the light Higgs boson has completed the particle spectrum of the standard model (SM) which can successfully describe many physical phenomena below or at around the electroweak (EW) scale, many fundamental issues remain unexplained today.
Following the symmetry principle and motivated by the issues of the SM, it is therefore appealing to consider a larger symmetry extending the SM at higher energy scale. This hypothesis implies the existence of an extended Higgs sector which is the strong motivations for going beyond the SM and makes the precision Higgs measurements becoming an important task. Discovery of more than one Higgs boson will be an indication of physics beyond the SM (BSM).

Since the Higgs sector in the SM suffers from the problems of naturalness and hierarchy, a light Higgs boson is technically unnatural.
It has been speculated that the Higgs could emerge as a composite pseudo-Goldstone boson (pGB) from extensions of the dynamical electroweak symmetry breaking patterns \cite{Composite1,Composite2,compositework1,compositework2,compositework3,compositework4,compositework5,compositework6}. The Elementary Goldstone Higgs (EGH) scenario \cite{EGH1,EGH2} take a complementary route by considering a fully perturbative SM extension in which an elementary pGB rather than a composite Goldstone Higgs naturally emerges from the precisely calculable dynamics. This model is the minimal underlying realization in terms of fundamental strongly coupled gauge theories supporting the global symmetry breaking pattern $SU(4)/Sp(4)$. The global symmetry is broken by the couplings with the electroweak (EW) gauge currents and SM Yukawa interactions. Under radiative corrections, these symmetry breaking will align the vacuum with respect to the EW symmetry, so the EW scale is radiatively induced. The embedding of the EW gauge sector is parameterized by the preferred electroweak alignment angle $\theta$. Its value is completely determined by the radiative corrections and the requirement that the model reproduces the phenomenological success of the Standard Model, which has been dynamically determined to be centered around $\theta\approx0.018$ \cite{EGH2}. The observed Higgs boson becomes a fundamental pseudo Nambu-Goldstone (PNG) boson. It can obtain a light mass through radiative corrections which could cause the symmetry breaking and explain the origin of the known fermion masses.


The template Higgs sector leading to the $SU(4) \rightarrow Sp(4)$ pattern of chiral symmetry breaking was first introduced in \cite{history1,history2,history3}. The relations of the EGH idea with unification scenario, the relaxation leptogenesis mechanism, and supersymmetry have been studied in references \cite{1511.01910,1601.07753} and \cite{susy}, respectively. The SM Higgs inflation has also been discussed in the context of the EGH model \cite{1611.04932}. The EGH model could produce a rich phenomenology because of the new scalar degrees of freedom.

Higgs boson pair production is well known for its sensitivity to the tri-Higgs coupling which provides a way to test the structure of Higgs potential and further EW symmetry breaking mechanism. Some of the relevant studies can be found in Refs. \cite{1212.5581,pphh1,pphh2,pphh3,pphh4,pphh5,pphh6,pphh7,pphh8,pphh9,pphh10,pphh11,eehh0,eehh1,eehh2,eehh3,eehh4,eehh5,eehh6,eehh7,eehh8,Yue}. The cross section of Higgs pair production is easily affected through modification in the top Yukawa coupling and the existence of new heavy scalars decaying into Higgs pairs. Thus, sizable production of Di-Higgs directly implies a new physics signature \cite{1212.5581}. Pair production of the EGH boson via gluon fusion at LHC has been studied in \cite{Yue}, which has shown that the resonant contribution of the heavy scalar is very small and the SM-like triangle diagram contribution is strongly suppressed. The total production cross section mainly comes from the box diagram contribution and its value can be enhanced with respect to the SM prediction.

The hadron collider has limitations to undertake precision studies due to the complexity of the final state and the smallness of the cross sections. The measurement of the Higgs trilinear self-coupling will be extremely challenging even at the high luminosity upgrade of the LHC. The $e^+e^-$ colliders, such as the International Linear Collider (ILC) \cite{ILC} and the Compact Linear Collider (CLIC) \cite{CLIC1,CLIC2}, operating at $\sqrt{s} =$ (500 GeV $\sim$ 3 TeV) are capable of measuring the Higgs cubic self couplings of the benchmark models with a good accuracy directly. A key physical goal for these different high-energy collider programs is try to probe the shape of the Higgs potential. Some of the recent studies of the Higgs pair searches in future colliders can be found in Refs. \cite{eehh0,eehh1,eehh2,eehh3,eehh4,eehh5,eehh6,eehh7,eehh8}. The relevant phenomenology has not been fully studied in the EGH model. In this work, we will focus on pair production of the EGH boson at the ILC and CLIC. There are three relevant modes: (i) Higgsstrahlung off a Z boson via $e^+e^-\rightarrow Zhh$, (ii) Vector Boson Fusion (VBF) via $e^+e^-\rightarrow \nu\bar{\nu}hh$ and (iii) associated production with top quarks via $e^+e^-\rightarrow t\bar{t}hh$. The process $e^+e^-\rightarrow e^+e^-hh $ has not been included as its cross section is about an order of magnitude smaller compared to $e^+e^-\rightarrow \nu\bar{\nu}hh$.

The rest of the paper is organized as follows. In Section II, we summarize the main features of the EGH model and show the relevant couplings. Pair production of the EGH boson via $e^+e^-$ collision is discussed in section III. Finally we present our conclusions in the last section.

\section{The main features of the EGH model}
The details of the EGH model have been studied in the original work Ref.~\cite{EGH1,EGH2}, including the particle spectrum, all couplings and interactions. Here we will briefly review the essential features of the model and focus our attention on the main features of the relevant parameters and couplings.

The Higgs sector is embedded into a chiral symmetry breaking of the pattern $SU(4)\rightarrow Sp(4)$.
The EW symmetry $SU(2)_L\times U(1)_Y$ is embedded in this $SU(4)$. A part of the $SU(4)/Sp(4)$ coset will be taken as the SM Higgs doublet.
And one Goldstone boson which acquires its mass via a radiatively-generated slight misalignment of the vacuum expectation value (VEV) will be identified as the Higgs boson. The Higgs sector strongly depends on the VEV $E_\theta$ which can be expressed in a general form as
\begin{eqnarray}
E_\theta=E_B\cos\theta +E_H\sin\theta  =-E_\theta^T,
\end{eqnarray}
where the misalignment angle $\theta \subset [0, \pi/2]$ and the independent VEVs $E_B$ and $E_H$ are given by
\begin{equation}
E_B=
\left(
\begin{array}{cc}
i\sigma_2 & 0 \\
0 & -i\sigma_2 \\
\end{array}
\right),\quad
E_H=
\left(
\begin{array}{cc}
0 & 1 \\
-1 & 0 \\
\end{array}
\right),
\end{equation}
with $\sigma_2$ being the second Pauli matrix. One can derive the allowed values of $\theta$ by considering both the quantum corrections and the requirement that the EGH model should be able to reproduce the SM's successes. It turns out that a small value of $\theta$ is preferred.

The above VEV alignment can be obtained by means of a scalar matrix $M$, which transforms as a two-index antisymmetric irreducible representation $\bf 6$ under $SU(4)$
\begin{eqnarray}
M=[\frac{\sigma+i\Theta}{2}+\sqrt{2}(i\Pi_i+ \widetilde{\Pi}_i) X_\theta^i]E_\theta .
\end{eqnarray}
Here $X_\theta^i$ (for $i=1,\ldots,5$) are the broken generators for the symmetry breaking $SU(4)\rightarrow Sp(4)$. (For details, see the Appendix A of Ref. \cite{EGH2}.)

In order to give the SM gauge sector, the $SU(2)_L\times U(1)_Y$ part of the chiral symmetry $SU(2)_L\times SU(2)_R\subset SU(4)$ is promoted to the gauge symmetry. Consequently, $M$ will couple to the EW gauge bosons via its covariant derivative
\begin{eqnarray}
D_\mu M=\partial_\mu M-i(G_\mu M+MG_\mu^T),
\end{eqnarray}
with
\begin{eqnarray}
G_\mu=gW_\mu^iT_L^i+g'B_\mu T_Y,
\end{eqnarray}
where $T_L^i$ (for $i=1,2,3$) and $T_Y=T_R^3$ are respectively the $SU(2)_L$ and hypercharge generators. So the kinetic term of $M$ appears as
\begin{eqnarray}
\mathcal{L}_{gauge}=\frac{1}{2}Tr[D_\mu M^\dagger D^\mu M].
\end{eqnarray}
Note that such a term explicitly breaks the global $SU(4)$ symmetry.

Two linear combinations $H_1$ and $H_2$ of the PNG bosons $\Pi_4$ and $\sigma$ are the mass eigenstates
\begin{equation}
\left(
\begin{array}{c}
H_1\\
H_2
\end{array}
\right)
=
\left(
\begin{array}{cc}
\cos\alpha & \sin\alpha \\
-\sin\alpha & \cos\alpha
\end{array}
\right)
\left(
\begin{array}{c}
\sigma \\
\Pi_4 \\
\end{array}
\right),
\end{equation}
with $\alpha$ (for $0<\alpha<\pi/2$) being the mixing angle. And the lighter one $H_1$ of them will be identified as the observed Higgs boson with a mass $m_h=125.09\pm0.24$ GeV \cite{mhiggs}.

The alignment of $E_\theta$ is dynamically determined by the renormalizability of the EGH model combined with the quantum corrections. It is found that \cite{EGH2} the preferred value of $\theta$ is $\theta=0.018_{-0.003}^{+0.004}$, corresponding to an $SU(4)$ spontaneous symmetry breaking scale of $f=13.9_{-2.1}^{+2.9}$ TeV as a result of the relation $f \sin\theta = \nu = 246$ GeV. This means that the EGH boson $H_1$ (which will be referred to as $h$) mainly consists of $\Pi_4$, while $H_2$ (which will be referred to as $H$) is mainly composed of $\sigma$.

The first three of the five PNG bosons $\Pi_i$ (for $i=1,\ldots,5$) become the longitudinal components of the EW gauge bosons $W$ and $Z$. On the other hand, Ref. \cite{EGH1} shows that $\Pi_5$ is a stable massive particle and provides a viable dark matter candidate. In the case of $M_{\Pi_5}=M_{DM}\geq m_h$, the EGH model is compatible with the experimental results.

The EGH model leads to the following SM normalised coupling strength of the scalars \cite{EGH1,EGH2}
\begin{eqnarray}
K_{h[H]}^F=\frac{g_{h[H]ff}}{g_{hff}^{SM}}=\sin(\alpha+\theta)[\cos(\alpha+\theta)],
\end{eqnarray}
\begin{eqnarray}
K_{h[H]}^V=\frac{g_{h[H]VV}}{g_{hVV}^{SM}}=\sin(\alpha+\theta)[\cos(\alpha+\theta)],
\end{eqnarray}
\begin{eqnarray}
K_{h[H]^2}^V=\frac{g_{h^2[H^2]VV}}{g_{h^2VV}^{SM}}=\sin^2(\alpha+\theta)[\cos^2(\alpha+\theta)],
\end{eqnarray}
\begin{eqnarray}
\mu_h=\frac{\lambda_{hhh}}{\lambda_{hhh}^{SM}}=\frac{M_\sigma^2\nu\cos\alpha}{fm_h^2},\quad
\mu_{Hh}=\frac{\lambda_{Hhh}}{\lambda_{hhh}^{SM}}=-\frac{M_\sigma^2\nu\sin\alpha}{3fm_h^2},
\end{eqnarray}
where $ff$ stands for all of the fermion pairs, $VV=WW$ and $ZZ$ and $\lambda_{hhh}^{SM}=3m_h^2/\nu$ is the SM trilinear self-coupling constant of the Higgs boson.
From the Ref.\cite{EGH2}, the Higgs self-coupling will increase with the vacuum alignment angle $\theta$.

The decay of the EGH bosons has been discussed in our previous work~\cite{Yue}.
The EGH boson $h$ has the same decay modes of the SM Higgs boson and its partial decay widths are universally shifted from the SM predictions by the factor $\sin^2(\alpha+\theta)$. The heavy scalar $H$ can decay to the SM gauge bosons and fermions with partial widths of
\begin{eqnarray}
\Gamma(H\rightarrow XX)=\cos^2(\alpha+\theta)\Gamma^{SM}(H\rightarrow XX),
\end{eqnarray}
where $\Gamma^{SM}(H\rightarrow XX)$ is the total decay width of the SM Higgs boson into the $XX$ final states evaluated at the $H$ mass $M_H$. At tree level, the partial decay widths for the processes $H\rightarrow hh$ can be written as
\begin{eqnarray}
\Gamma(H\rightarrow hh)=\frac{M_\sigma^4\sin^2\alpha}{32\pi f^2M_H}\sqrt{1-\frac{4m_h^2}{M_H^2}}.
\end{eqnarray}

Considering the heavy $H$ mainly from the scalar $\sigma$, we take $M_H\approx M_\sigma$ and assume its values in the range of 500 GeV $\sim$ 3 TeV.
In the following section, we will use these relations to consider possible pair production of the EGH boson, and focus our attention on the corrections to the SM prediction via $e^+e^-$ collision.

\section{Pair production of the EGH boson}
Pair production of the EGH boson $h$ in future linear electron positron colliders is mainly induced by three sources: (i) $Z$ strahlung process ($e^+e^-\rightarrow Z^{*}\rightarrow Zhh$), (ii) $W$ boson fusion mechanism ($e^+e^-\rightarrow\nu\bar{\nu}hh$) and (iii) associated production with top quarks ($e^+e^-\rightarrow t\bar{t}hh$). These production modes are useful to extract the Higgs trilinear self-coupling via (i) and (ii) plus the top-Yukawa coupling via (iii).


\begin{figure}
\begin{center}
\includegraphics [scale=0.9] {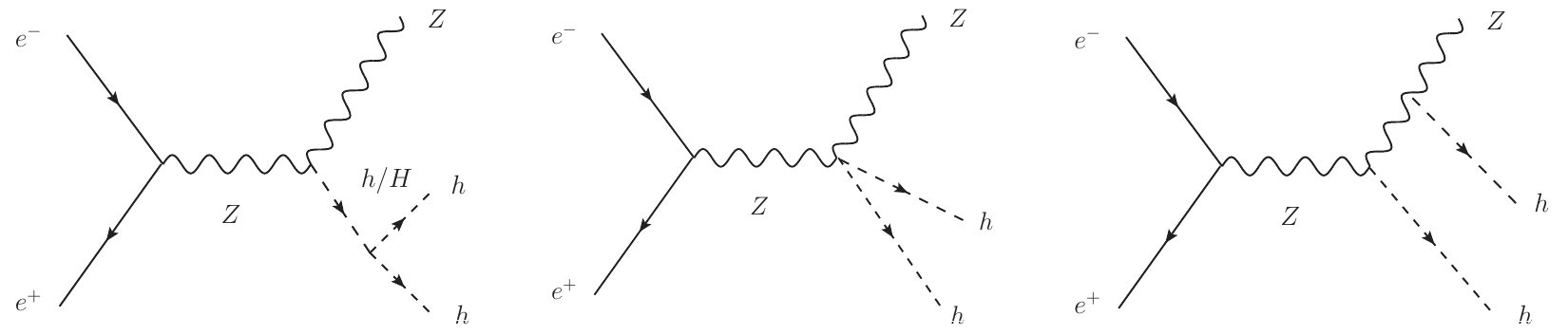}
\caption{Representative Feynman diagram for the $e^+e^-\rightarrow Zhh$ process.} \label{zhh}
\end{center}
\end{figure}
\begin{figure}
\begin{center}
\includegraphics [scale=0.35] {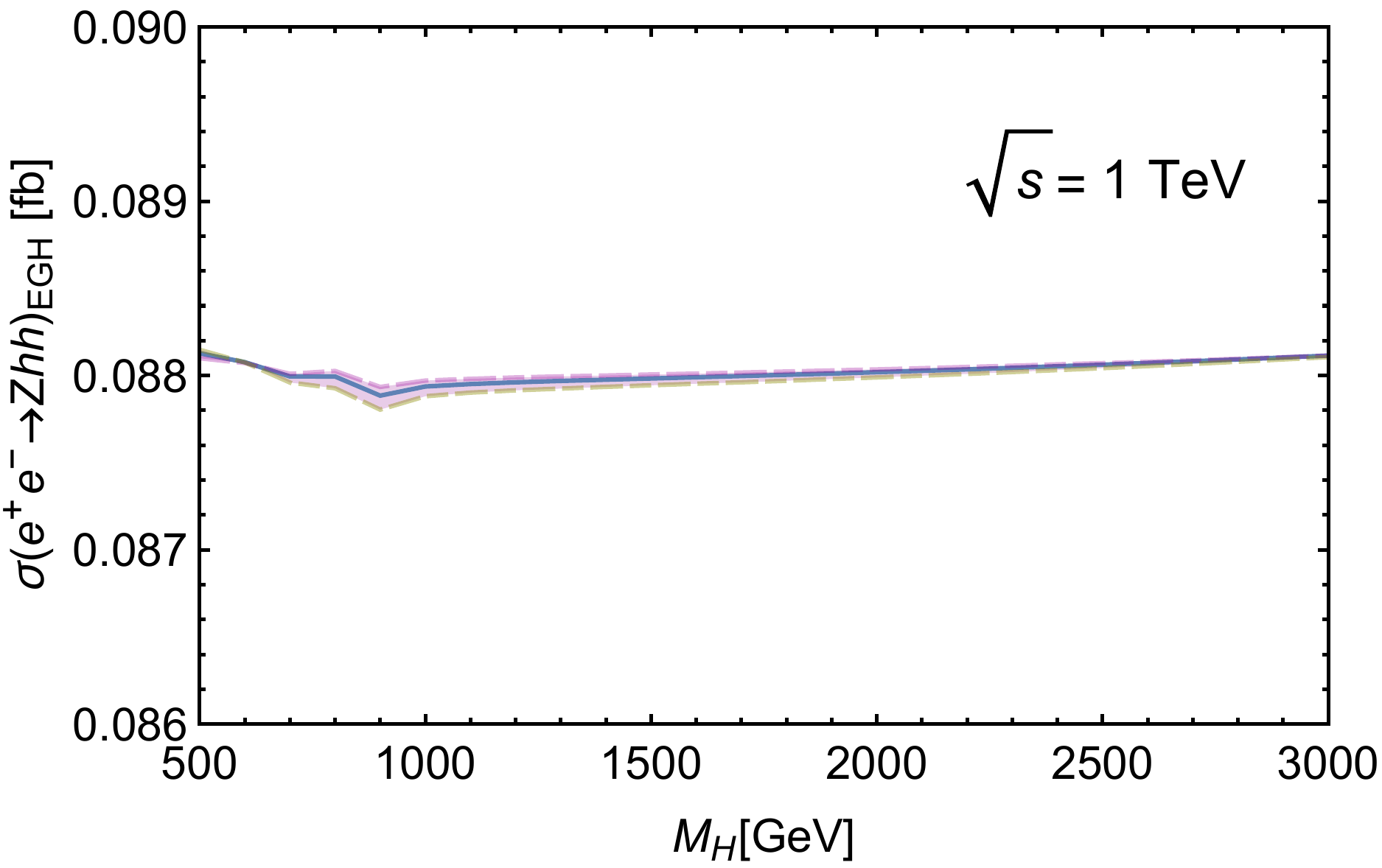}
\includegraphics [scale=0.35] {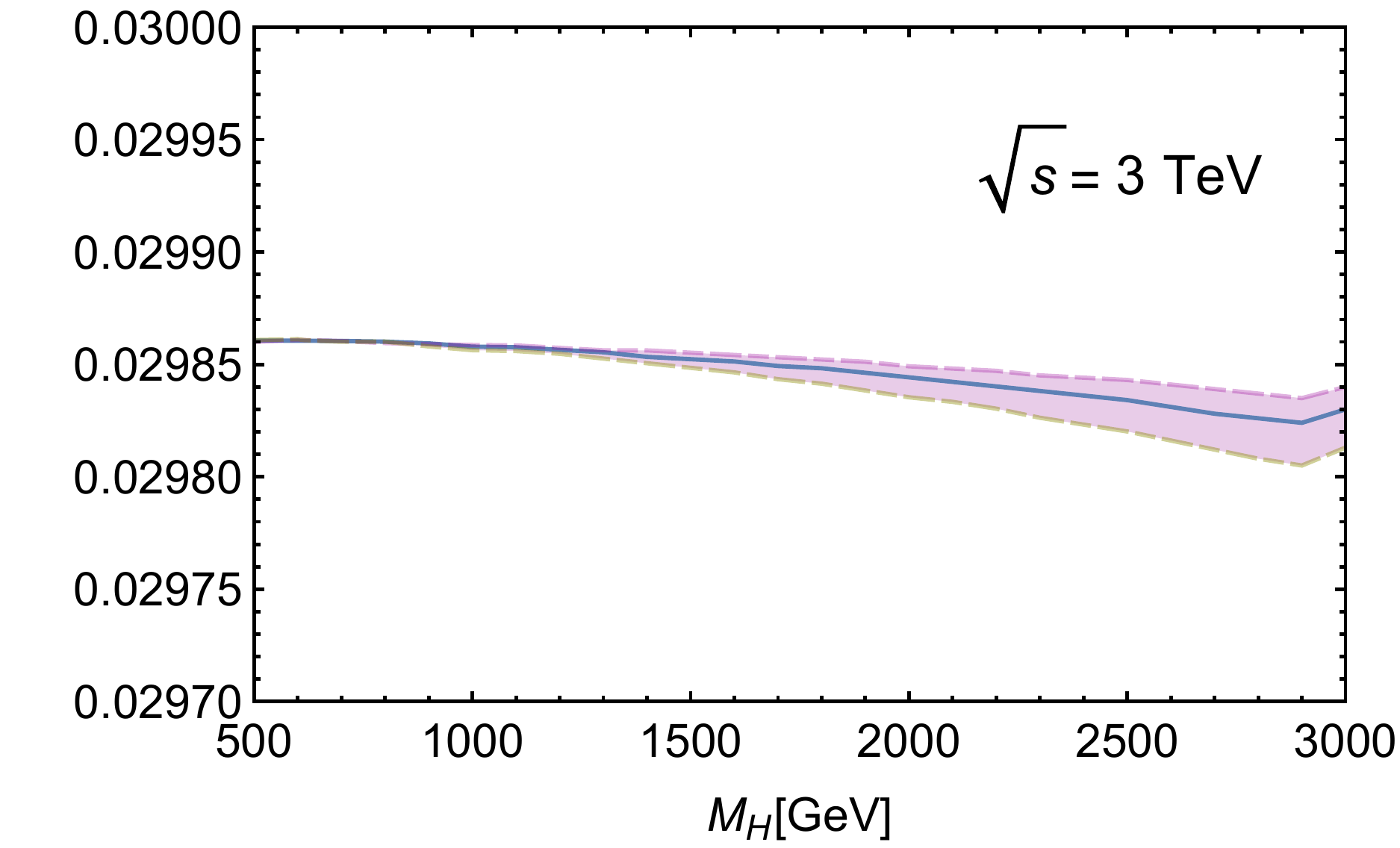}
\caption{The cross section of $e^+e^-\rightarrow Zhh$ as a function of the mass $M_H$ for the mixing angle $\alpha=1.57$ and the vacuum alignment angle $\theta=0.018$ within the statistical error on $\theta$ (the pink region).  In the left panel, the center of mass (c.m.) energy is 1 TeV for the ILC. In the right panel, we set the collision energy at 3 TeV for the CLIC.} \label{cs:zhh}
\end{center}
\end{figure}

First of all, let us discuss the $Zhh$ production mode. The representative Feynman diagrams are shown in Fig.\ref{zhh}.
The $e^+e^-\rightarrow Zhh$ process in future $e^{+}e^{-}$ colliders plays an important role in measurement of the $hhh$ coupling constant for light Higgs bosons due to the simple kinematical structure. But three body final states of $hhZ$ requires relatively larger collision energy. As a result, the $s$-channel nature of the process may decrease the cross section.

We take the preferred values $\theta=0.018_{-0.003}^{+0.004}$ and $\alpha=1.57$ given by Ref.\cite{EGH2} which satisfy the current LHC experimental constraint. In Fig.\ref{cs:zhh}, the cross sections of the $Z$ strahlung process are evaluated as a function of $M_H$ by Madgraph5/aMC@NLO \cite{mg5amc}. According to the integrated luminosities of $1~ab^{-1}$ and $2~ab^{-1}$ provided by 1 TeV ILC and 3 TeV CLIC \cite{ILCTDR1,ILCTDR2}, the number of event will not be influenced by the mass parameter $M_H$ in the range of 500 GeV $\sim$ 3 TeV. In the EGH model, the trilinear self-coupling of the SM-like Higgs boson with respect to the SM one is strongly suppressed, which causes less contribution to cross section of the first Feynman diagram in Fig.\ref{zhh}. Therefore, the cross section of the $Z$ strahlung process is slightly influenced by the mass parameter $M_H$.

The contributions of top quark and weak gauge bosons may induce a deviation of the cross section by loop-induced couplings. Such deviation depends on the embedding angle $\theta$, which is completely determined by the radiative corrections and the requirement that the model reproduces the phenomenological success of the SM. As discussed in Ref. \cite{EGH2}, rather large $\theta$ is allowed in some choices of model parameters. In such cases, the contributions of top quark and weak gauge bosons should not be ignored. But with the sufficiently small $\theta$, this kind of contributions to the cross section could be negligible. Furthermore, the evolution of the Higgs self-couplings in the SM decreases with the increase of scale because of the RG running \cite{HiggsRG}. In the EGH model, the evolution of the scalar self-couplings has similar behavior with the SM ones. At the scales relevant for the considered processes, the trilinear scalar self-couplings increase due to the RG running effects. However, these corrections can be ignored. On one hand, from Fig.2, we find that the cross section of the $Z$ strahlung process is insensitive to $M_H$, so the RG running on $M_\sigma$ which equals to $M_H$ has little effect on $\lambda_{Hhh}$. On the other hand, the value of $\lambda_{hhh}$ is very small with $\alpha=1.57$ and it will be little affected by RG running. Therefore, the modifications to the results are negligible.

\begin{figure}
\begin{center}
\includegraphics [scale=0.28] {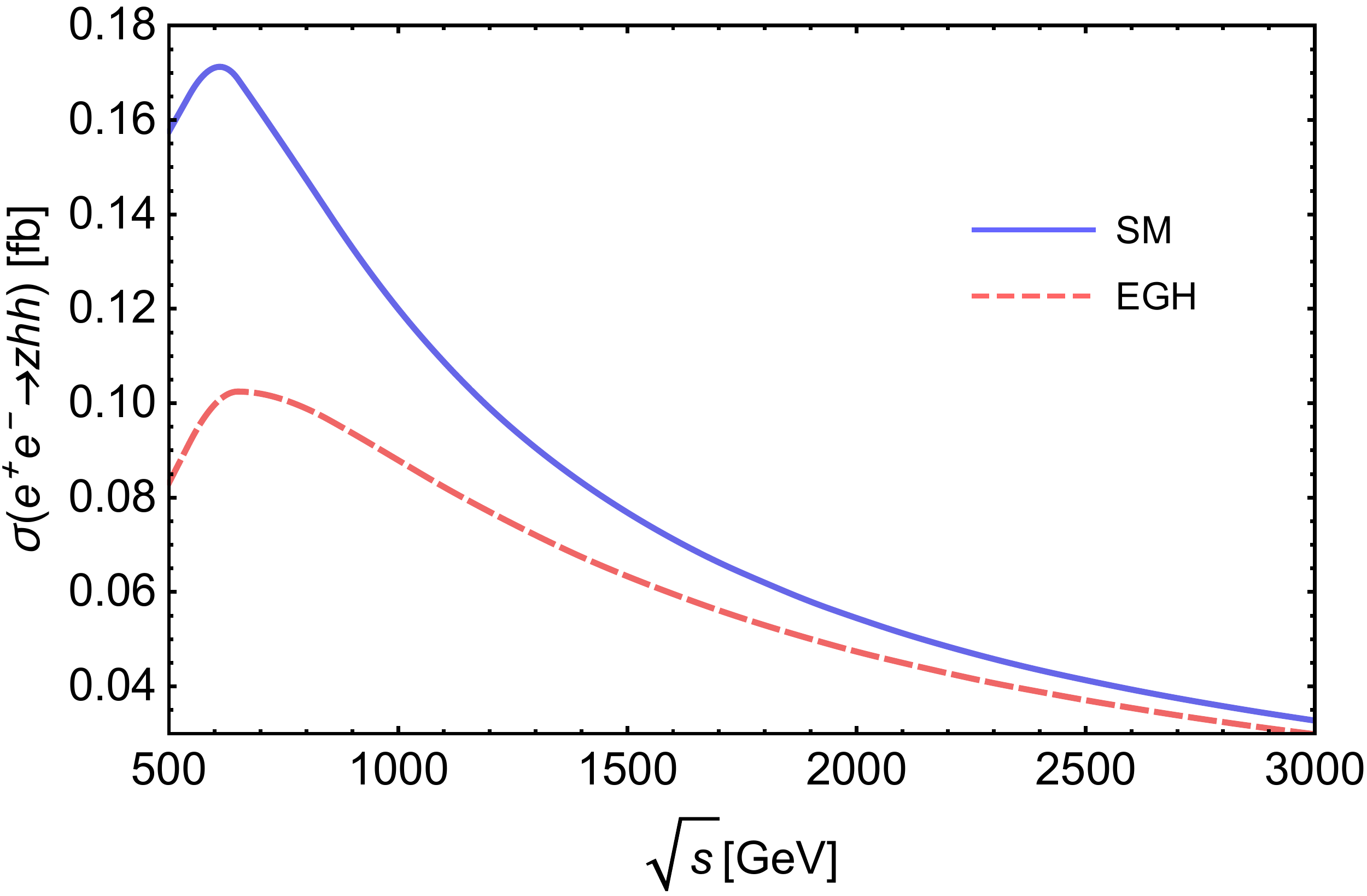}
\includegraphics [scale=0.27] {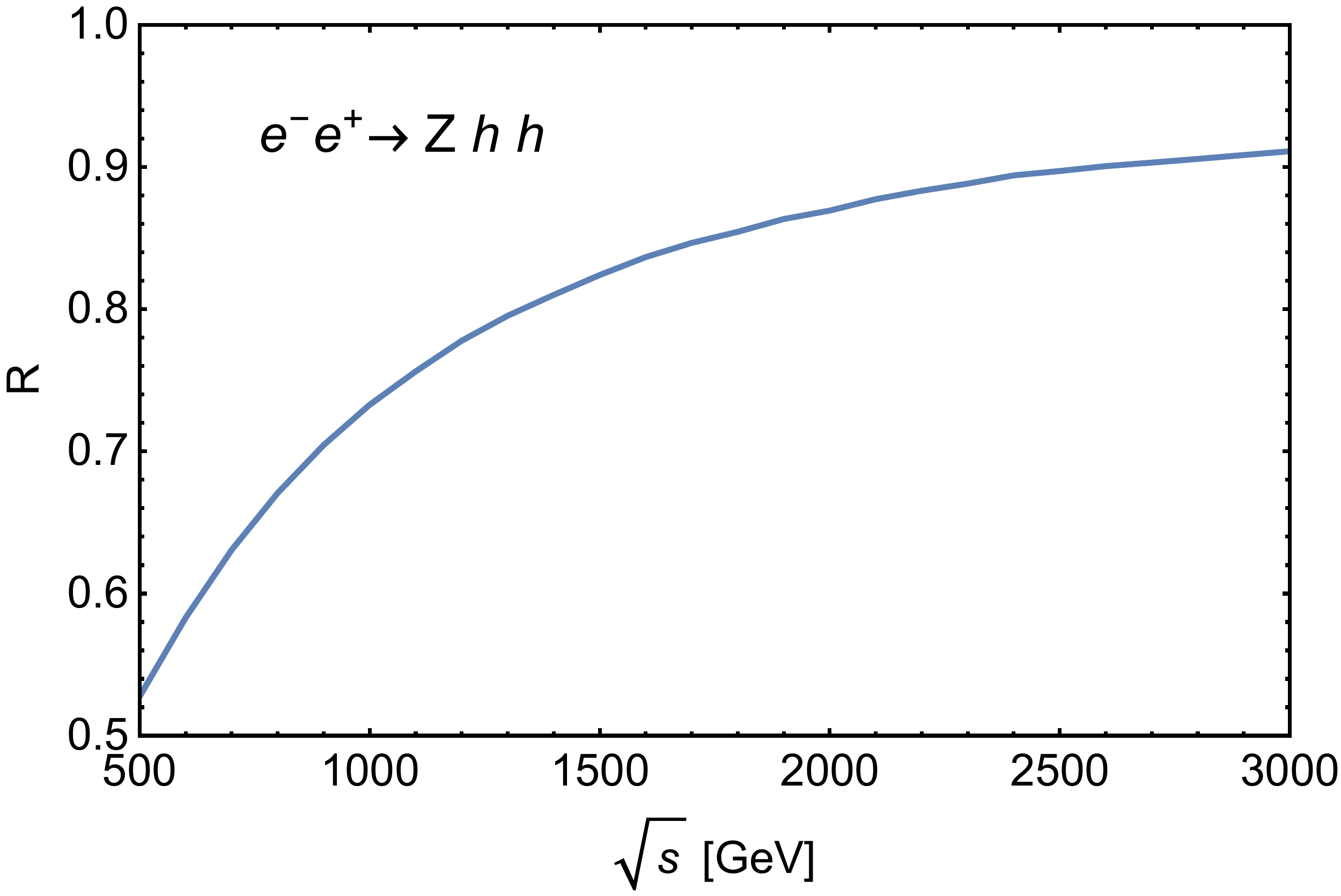}
\caption{The cross section and the ratio $R=\sigma^{EGH}/\sigma^{SM}$ for the $e^+e^-\rightarrow Zhh$ process as a function of c.m. energy for $M_H = 1$ TeV, $\alpha=1.57$ and $\theta=0.018$.} \label{R:zhh}
\end{center}
\end{figure}
For the $e^+e^-\rightarrow Zhh$ process, the cross section is smaller than that in the SM. Our numerical results show that it is indeed this case. Fig.\ref{R:zhh} shows the ratio $R=\sigma^{\rm{EGH}}/\sigma^{\rm{SM}}$ as a function of the center of mass (c.m.) energy $\sqrt{s}$ with $M_H = 1$ TeV, $\alpha=1.57$ and $\theta=0.018$.
The contributions of the first Feynman diagram and its interference with the second and third diagrams reduce faster than the SM ones with the c.m. energy increasing. The cross sections of the second and third diagrams in Fig.1 are similar to the SM ones. Thus, for the $Z$ strahlung process, the biggest deviation shows at the lower energy in Fig.3.
The correction value $\delta R=(\sigma^{EGH}-\sigma^{SM})/\sigma^{SM}$ is about $-9\% \sim -47\%$ in the range of $\sqrt{s}=$ 500 GeV $\sim$ 3 TeV ($-27\%$ for 1 TeV). The production cross section $e^+e^-\rightarrow Zhh$ is measured with accuracy of 42.7 \% (23.7 \%) at the ILC with $\sqrt{s}$ = 500 GeV and the integrated luminosity of 500 fb$^{-1}$(1600 fb$^{-1}$)\cite{1310.0763}. According to these expected accuracy of measurements, such deviation might be observed by the high-luminosity at the ILC.

We compare the results with the previous work \cite{eehh4} which discussed the $e^+e^-\rightarrow Zhh$ process in three different minimal composite Higgs models (MCHMs). In the range of $\sqrt{s}$ = 400 GeV $\sim$ 1.5 TeV, $\sigma^{\rm{MCHM}}/\sigma^{\rm{SM}}$ is supposed to be about 0.69(0.38) $\sim$ 0.75(0.54) with the compositeness parameter $\xi$ = 0.1(0.2). The results show that the deviation from the SM in the EGH model is more obvious than ones in the MCHMs within the lower energy range.


\begin{figure}
\begin{center}
\includegraphics [scale=1.1] {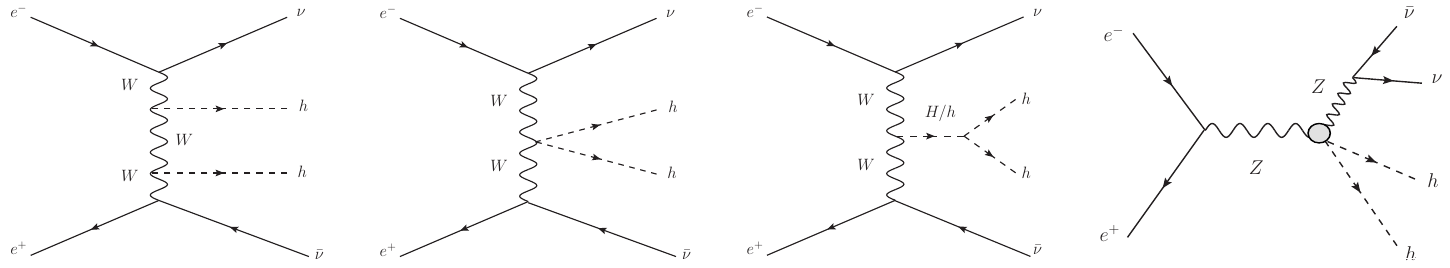}
\caption{Representative Feynman diagram for the $e^+e^-\rightarrow\nu\bar{\nu}hh$ process. The last diagram corresponds to the process $e^+e^-\rightarrow Zhh\rightarrow\nu\bar{\nu}hh$ (see Fig.1). } \label{vvhh}
\end{center}
\end{figure}

Next, we discuss double-$h$ production via the $W$ boson fusion mechanism. The representative Feynman diagrams are shown in Fig.\ref{vvhh}. The $Z$ strahlung topologies (last diagram in Fig.\ref{vvhh}) play a subdominant role due to the tiny branching ratio of $Z\rightarrow\nu\bar{\nu}$. The best energy to investigate the Higgsstrahlung production $e^+e^-\rightarrow Zhh$ is around 600 GeV, however the $e^+e^-\rightarrow\nu\bar{\nu}hh$ at 600 GeV is very small. The cross section of $WW$-fusion production will increase with the energy because of the $t$-channel enhancement of $W^+W^-\rightarrow hh$ subprocess. At TeV energies, the  cross section of fusion processes dominates over that from $e^+e^-\rightarrow Zhh$ and would give significant event rates at high luminosity $e^+e^-$ colliders \cite{HanTao}, thus the $e^+e^-\rightarrow\nu\bar{\nu}hh$ process becomes more important for higher $\sqrt{s}$. This is one essential motivation to go to higher energy after running at 500 GeV.

\begin{figure}
\begin{center}
\includegraphics [scale=0.35] {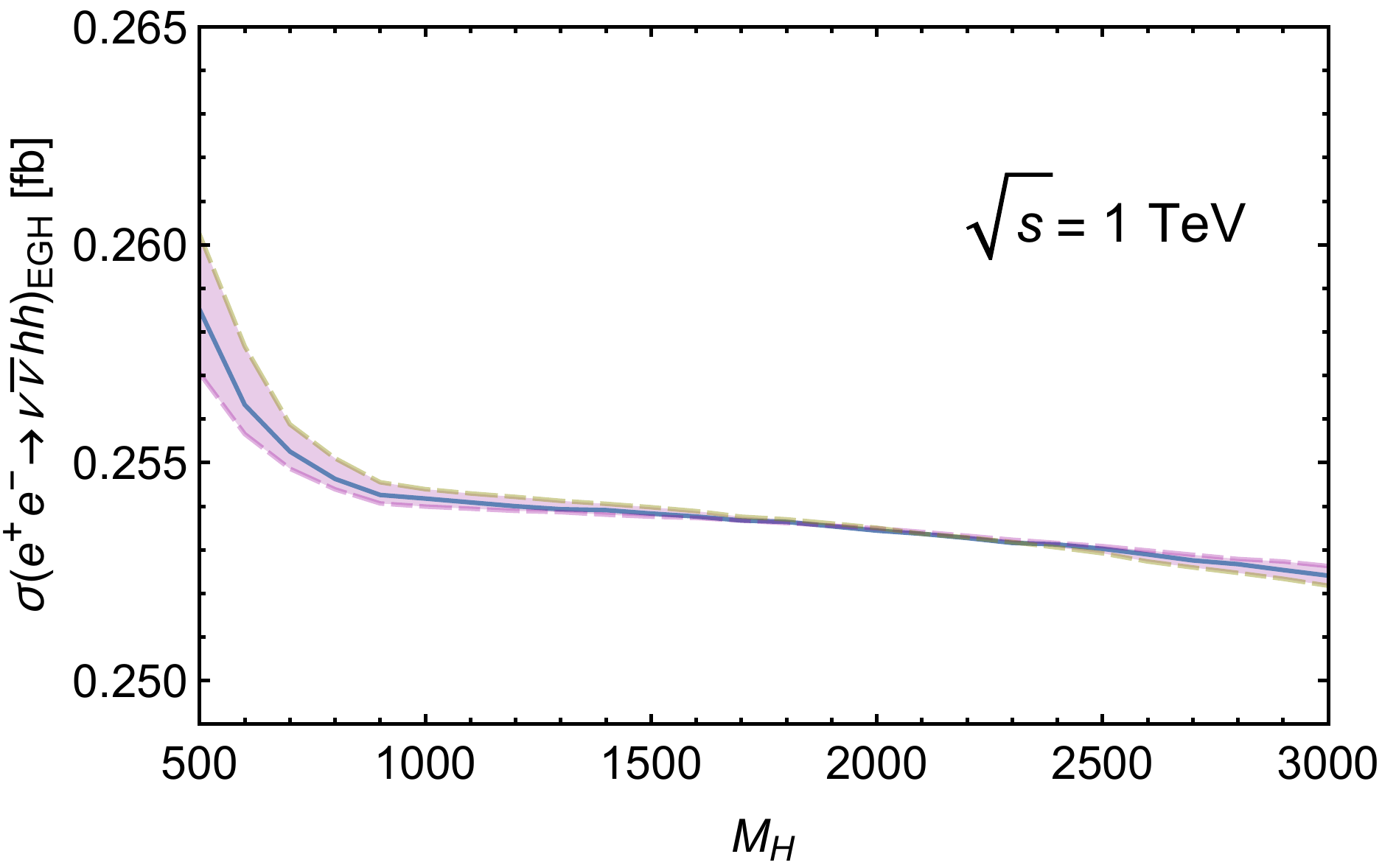}
\includegraphics [scale=0.35] {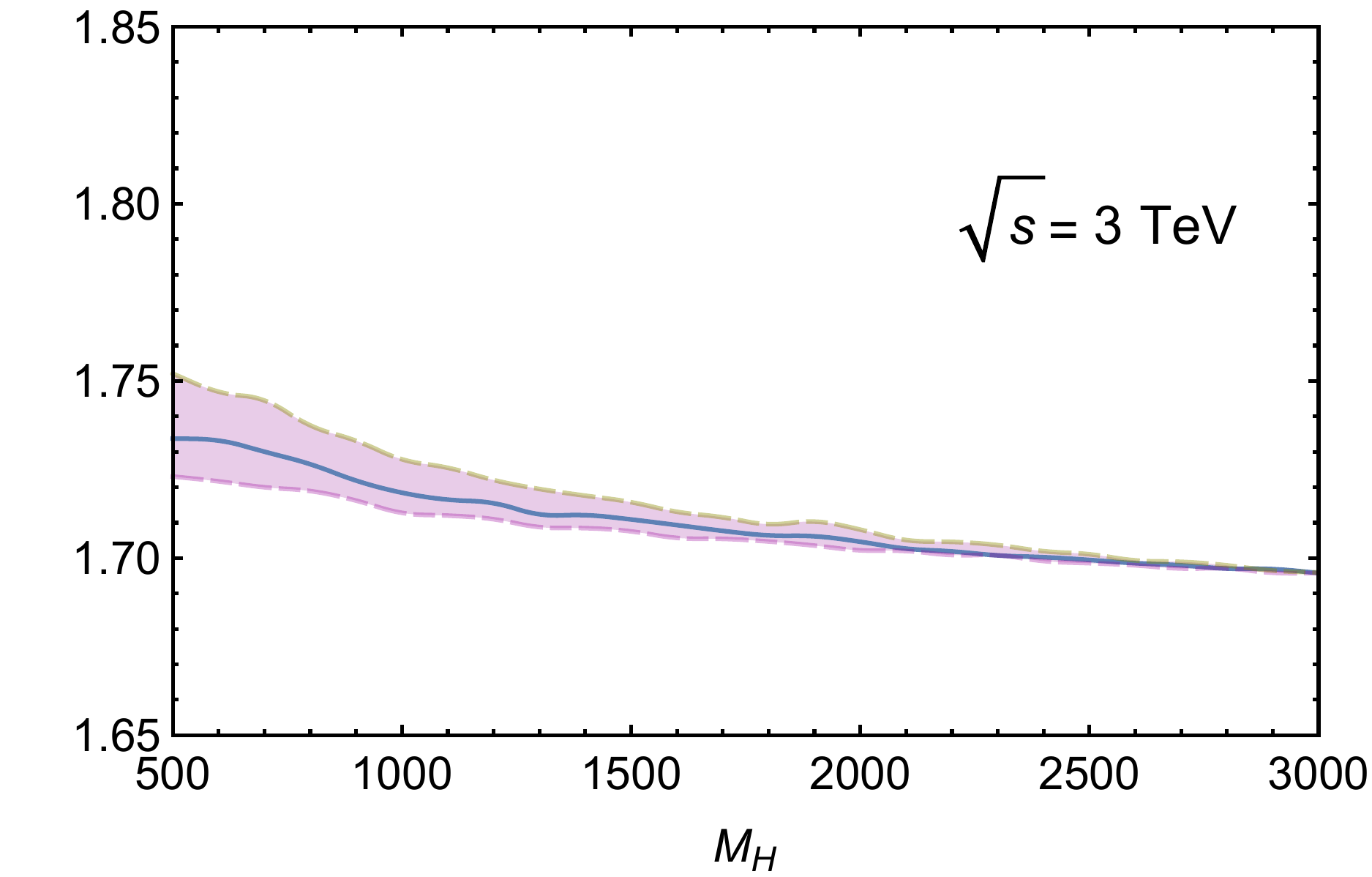}
\caption{The cross section of $e^+e^-\rightarrow\nu\bar{\nu}hh$ at the ILC (left) and the CLIC (right) as a function of the mass $M_H$ for $\alpha=1.57$ and $\theta=0.018_{-0.003}^{+0.004}$.} \label{cs:vvhh}
\end{center}
\end{figure}

We evaluate the production rate for $e^+e^-\rightarrow\nu\bar{\nu}hh$ in Fig.\ref{cs:vvhh}. It can be seen that the cross section is mostly insensitive to the mass parameter. However, the $M_H$ dependence in this process is rather large compared to the other channels especially in the range of 500 GeV $<M_H<$ 1000 GeV. Because this process is equal to direct production of the $WW\rightarrow hh$, which gives rise to a better sensitivity to the mass parameter $M_H$ than that in the secondary decay of $e^+e^-\rightarrow ZH$.

\begin{figure}
\begin{center}
\includegraphics [scale=0.28] {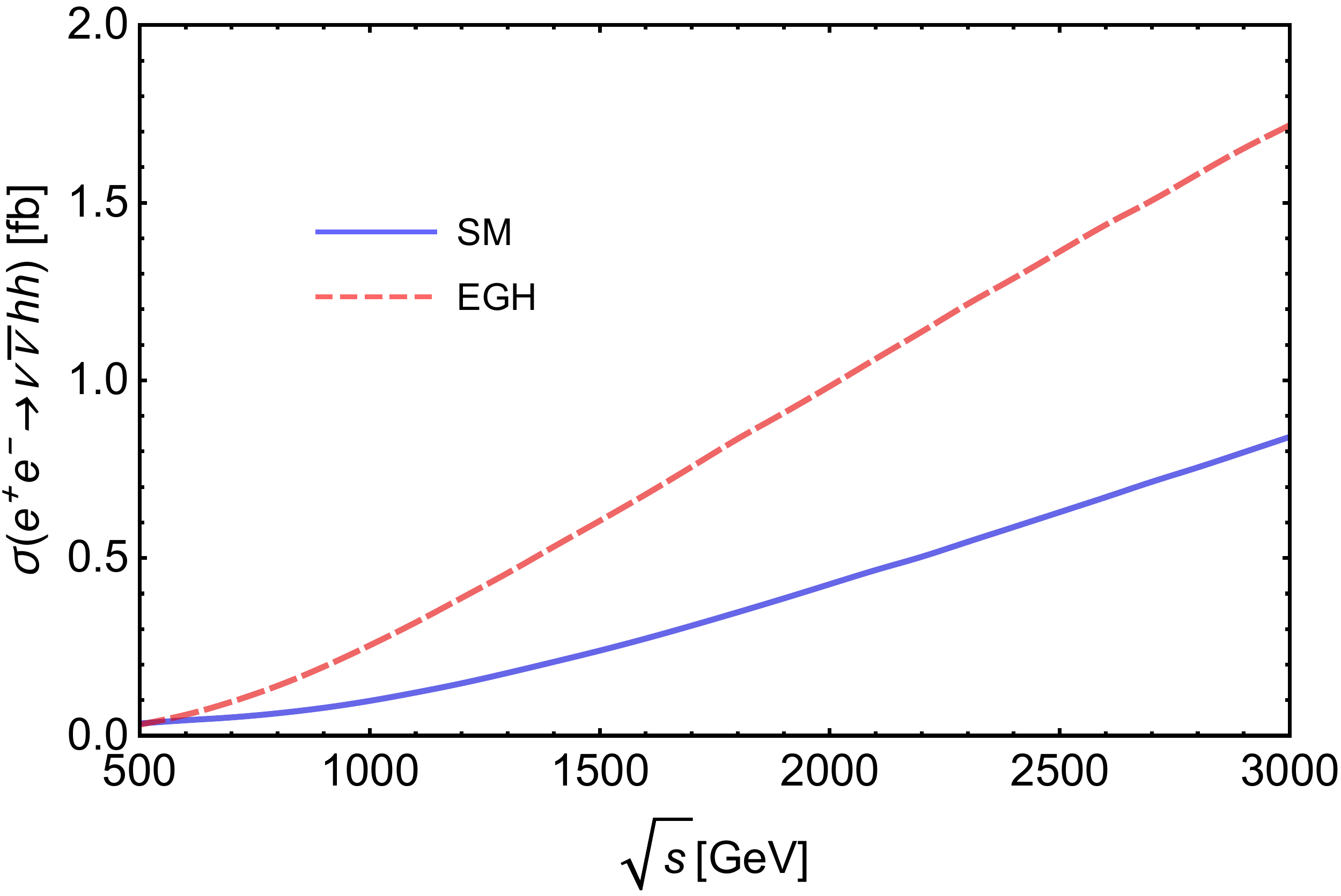}
\includegraphics [scale=0.28]  {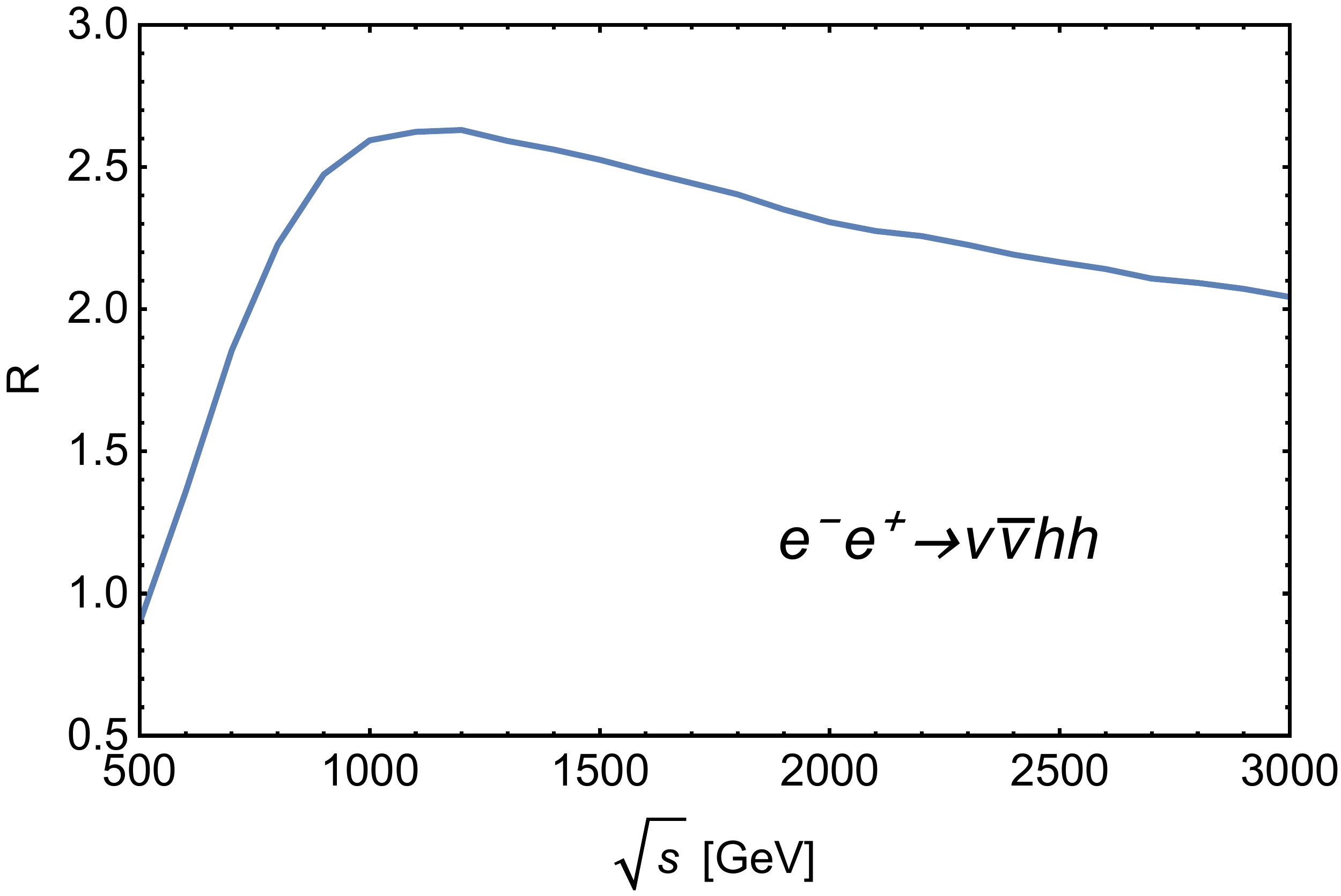}
\caption{The cross section and the ratio $R=\sigma^{EGH}/\sigma^{SM}$ for the $e^+e^-\rightarrow \nu\bar{\nu}hh$ process as a function of c.m. energy. We take $M_H = 1$ TeV, $\alpha=1.57$ and $\theta=0.018$.} \label{R:vvhh}
\end{center}
\end{figure}

As the benchmark point, we take $M_H = 1$ TeV, $\alpha=1.57$ and $\theta=0.018$ to show the ratio $R$ as a function of $\sqrt{s}$ in Fig.\ref{R:vvhh}. One can see from this figure that the production rate for pair production of the SM-like Higgs via the $W$ boson fusion in the EGH model is significantly larger than the SM prediction. For $\alpha=1.57$, $\theta=0.018$ and 500 GeV $\leq \sqrt{s}\leq$ 3 TeV, the value of the ratio $\delta R$ is in the range of $-10\% \sim 163\%$ with the peak around 1.2 TeV. It is expected that the production cross section of the $e^+e^-\rightarrow \nu\bar{\nu}hh$ process will be measured with the accuracy of 26.3 \% (16.7 \%) at the ILC with $\sqrt{s}$ = 1 TeV and the integrated luminosity of $1~ab^{-1}$ ($2.5~ab^{-1}$)\cite{1310.0763}. For CLIC, the expected uncertainties for $1.5~ab^{-1}$ at $\sqrt{s}$ = 1.4 TeV and $2~ab^{-1}$ at $\sqrt{s}$ = 3 TeV are 32\% and 16\%. Hence such deviation should be measured at the ILC and CLIC.

For the $Z$ strahlung and $W$ boson fusion processes, there is only one diagram containing the $Hhh$ vertex. Thus, these production cross sections of the corresponding ones are proportional to the squared Higgs trilinear coupling $\lambda^2_{Hhh}$.
In EGH model, $\lambda_{Hhh}$ depends on two parameters, $\alpha$ and $M_\sigma$, as we can see in Eq.(10). For the benchmark point $\alpha=1.57$, $\theta=0.018_{-0.003}^{+0.004}$ and $M_H\approx M_\sigma$, the weak gauge couplings $\kappa^V$ are fixed, and the trilinear Higgs coupling only depends on one parameter $M_H$.  It is possible for $W$ boson fusion processes to obtain an expected sensitivity on the trilinear Higgs coupling via measuring the cross section at the fixed $\kappa^V$, but it is challenging for $Z$ strahlung processes at $e^+e^-$ colliders, because they have little sensistivity to $M_H$. Furthermore, in most of the range of the c.m. energies, the correction effect of $e^+e^-\rightarrow\nu\bar{\nu}hh$ process is positive, which is opposite to the $Z$ strahlung. Therefore, the VBF process has an advantage to obtain better sensitivities to measure the trilinear self-coupling of the EGH boson.

We also compare the ratio $R$ of the $e^+e^-\rightarrow \nu\bar{\nu}hh$ process with that in MCHMs \cite{eehh4}.
In the range of $\sqrt{s}$ = 400 GeV $\sim$ 1.5 TeV, $\sigma^{\rm{MCHM}}/\sigma^{\rm{SM}}$ is supposed to be about 0.5(0.7) $\sim$ 2.0(1.4) with the compositeness parameter $\xi$ = 0.1(0.2).  The expected $R$ of the $e^+e^-\rightarrow \nu\bar{\nu}hh$ process in EGH model is larger than that in MCHMs for the same value of $\sqrt{s}$.


\begin{figure}
\begin{center}
\includegraphics [scale=1] {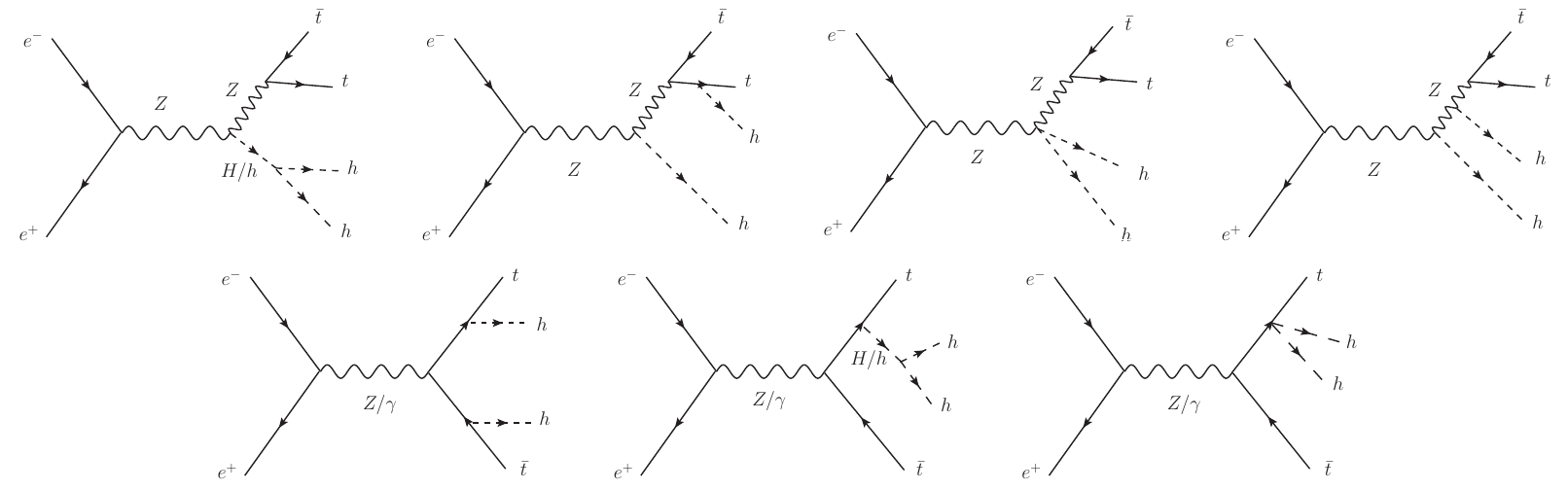}
\caption{Representative Feynman diagram for the $e^+e^-\rightarrow t\bar{t}hh$ process. } \label{tthh}
\end{center}
\end{figure}

Finally, we investigate the $t\bar{t}hh$ production process, for which the representative Feynman diagrams are shown in Fig.\ref{tthh}. Differently from the previous two double-$h$ production modes, this process could measure not only the trilinear couplings but also top-Yukawa couplings. In addition, this mode is useful to extract $\tan\beta$ in context of 2HDM either elementary or composite \cite{eehh6}.

\begin{figure}
\begin{center}
\includegraphics [scale=0.38] {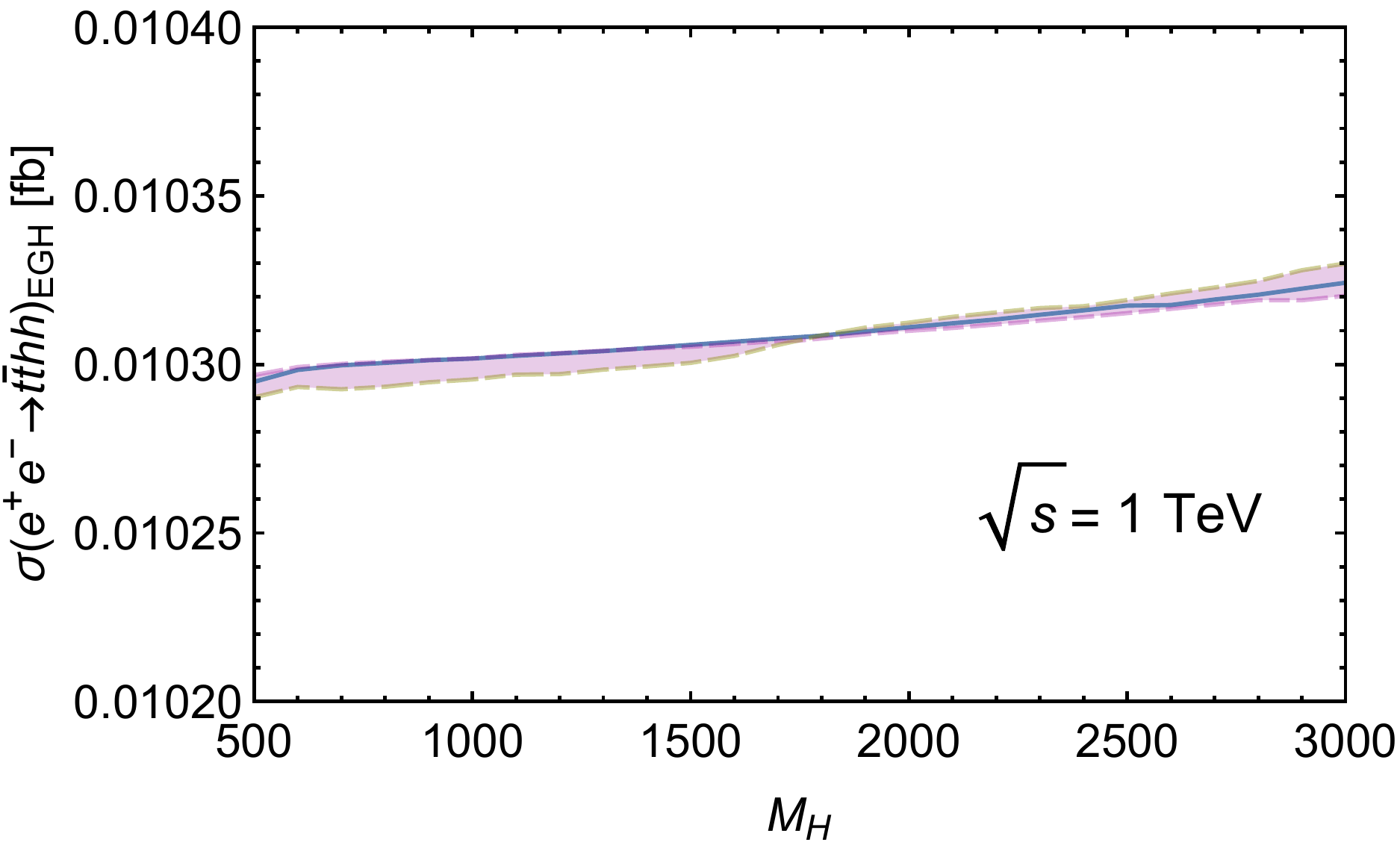}
\includegraphics [scale=0.38] {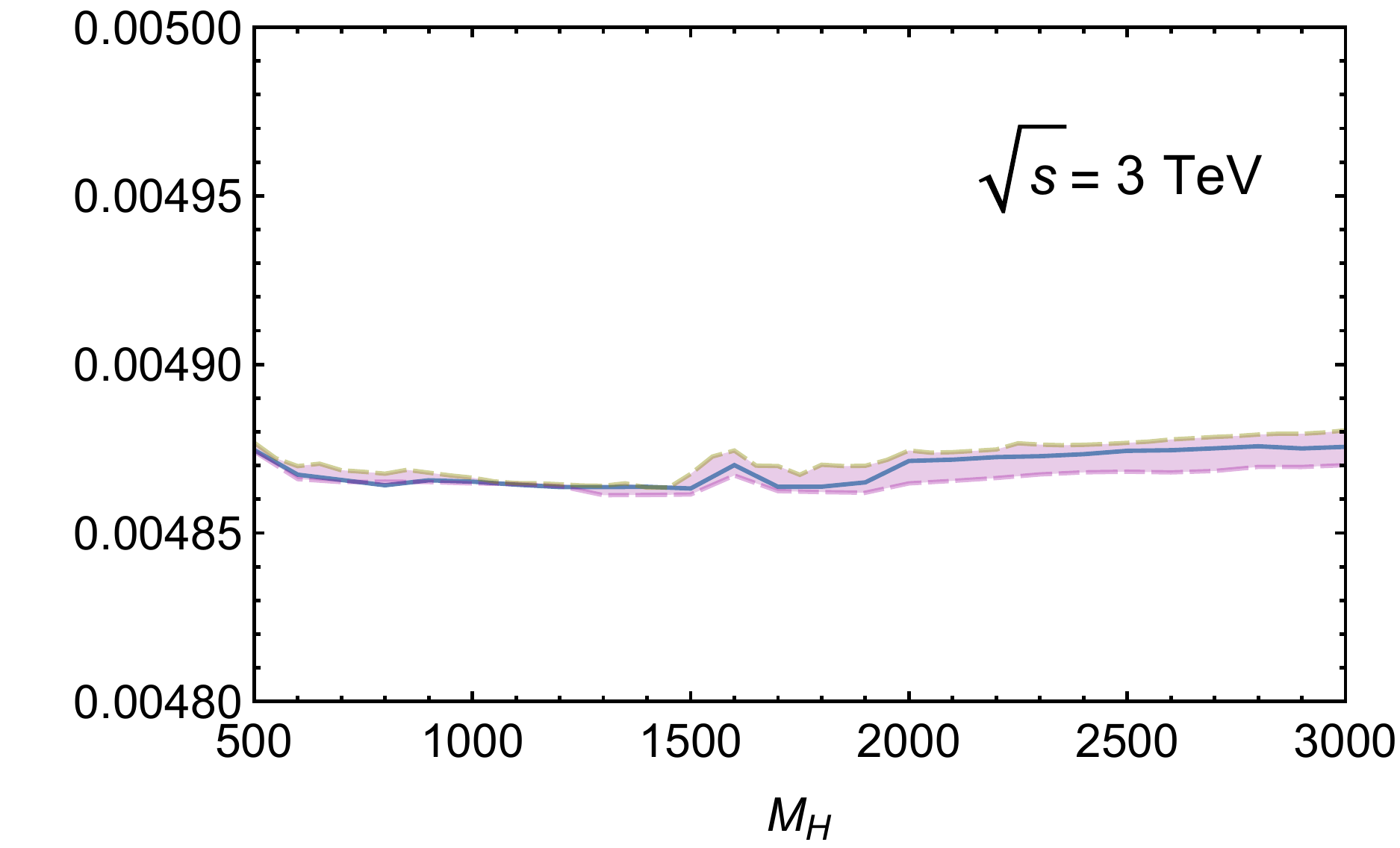}
\caption{The cross section of $e^+e^-\rightarrow t\bar{t}hh$ at the ILC (left) and the CLIC (right) as a function of the mass $M_H$ for $\alpha=1.57$ and $\theta=0.018_{-0.003}^{+0.004}$.} \label{cs:tthh}
\end{center}
\end{figure}

\begin{figure}
\begin{center}
\includegraphics [scale=0.3] {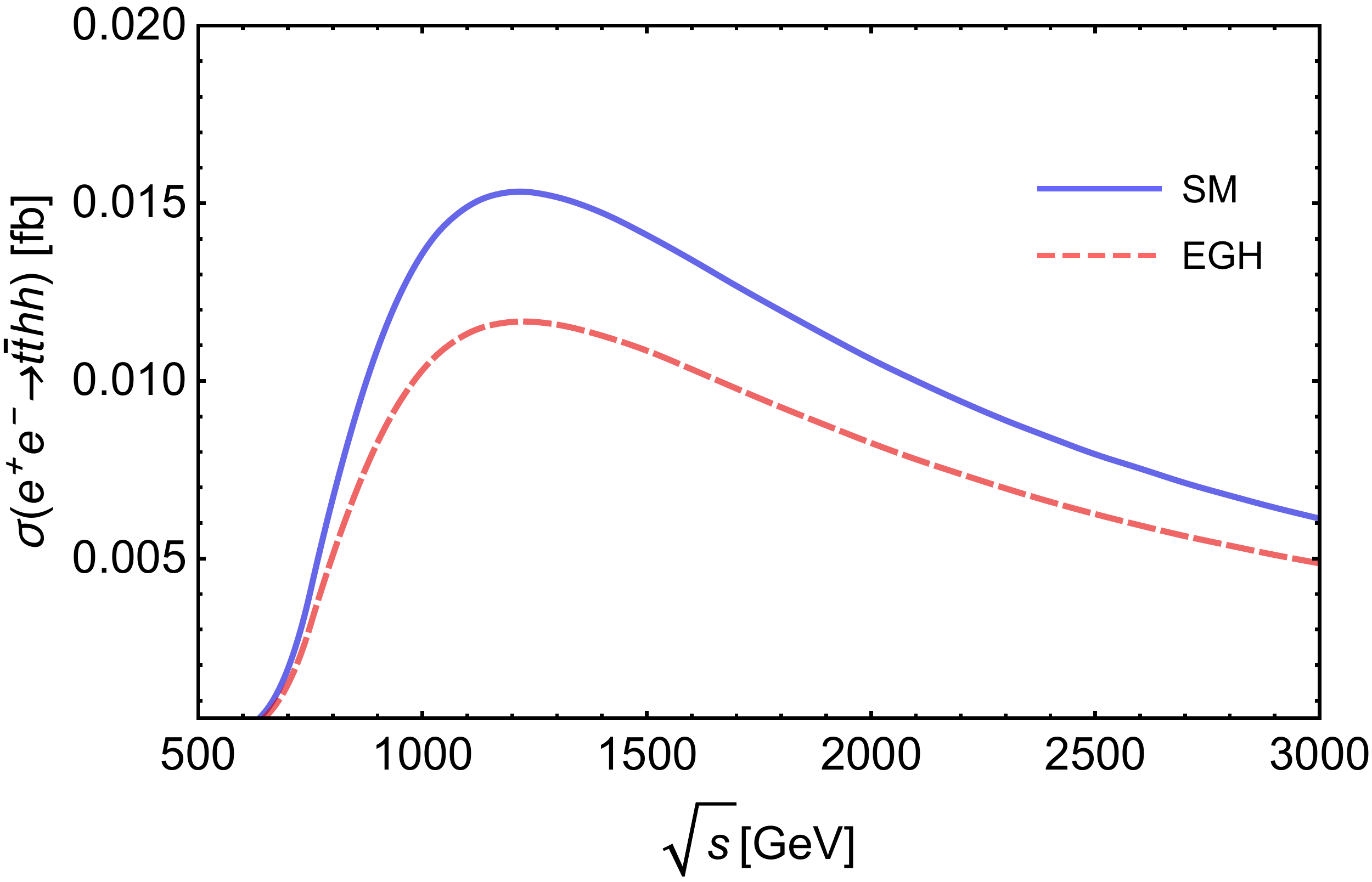}
\includegraphics [scale=0.29] {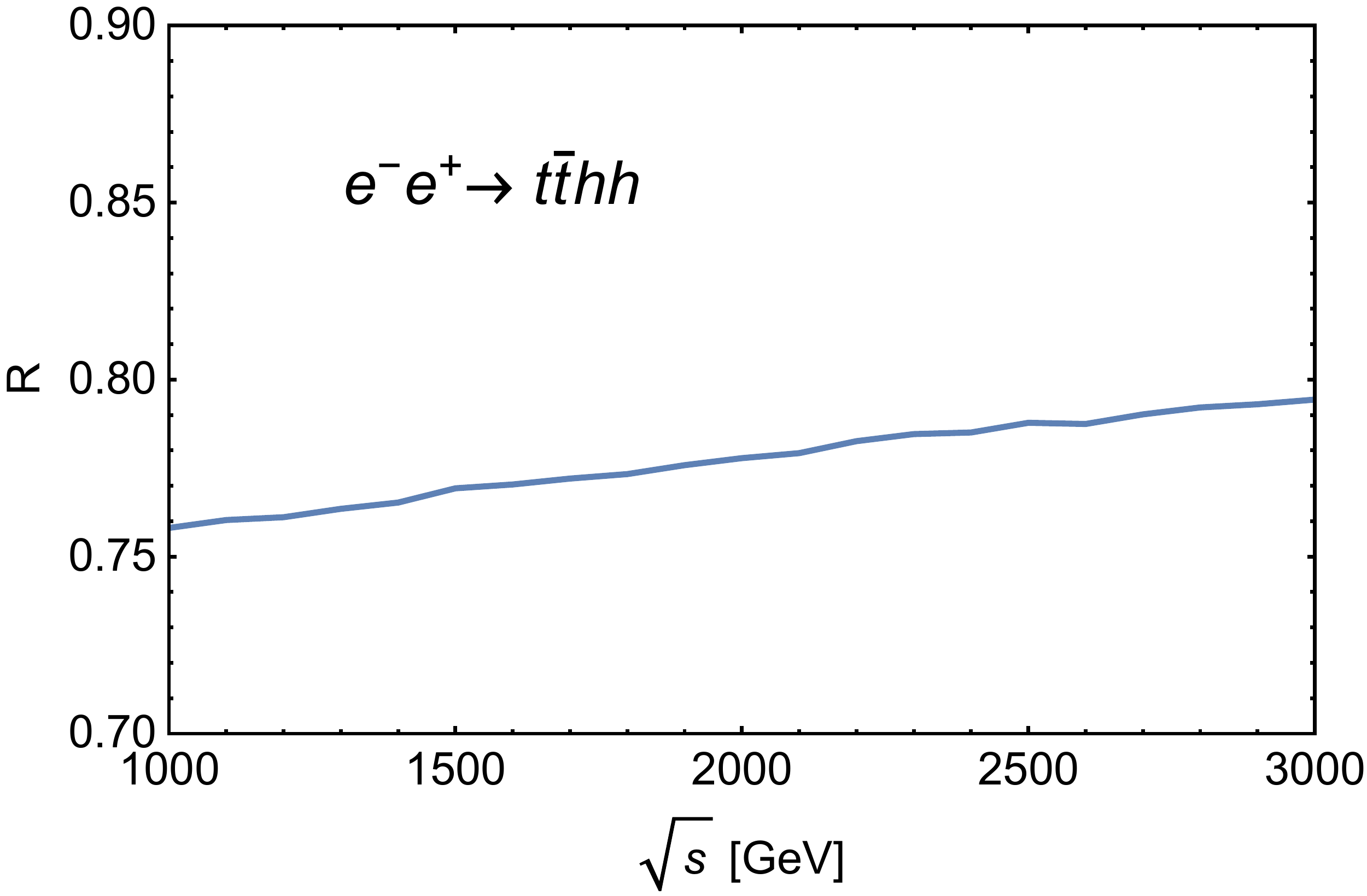}
\caption{The cross section and the ratio $R=\sigma^{EGH}/\sigma^{SM}$ for the $e^+e^-\rightarrow t\bar{t}hh$ process as a function of c.m. energy. We take $M_H = 1$ TeV, $\alpha=1.57$ and $\theta=0.018$.} \label{R:tthh}
\end{center}
\end{figure}

In Fig.\ref{cs:tthh}, we plot the cross section of $e^+e^-\rightarrow t\bar{t}hh$ as a function of the mass parameter $M_H$.
It can be seen that the production cross section is very small and less sensitive to the mass parameter $M_H$. This behaviour can be explained as fellows. Besides the trilinear self-coupling of $h$ is strongly suppressed, the couplings of the heavy scalar $H$ to the SM fermion pairs are also strongly suppressed by the factor $\cos(\alpha+\theta)$ for $\alpha=1.57$ and $\theta=0.018_{-0.003}^{+0.004}$. Therefore, its resonant contribution to Di-Higgs production cross section is negligible. We plot the ratio $R$ as a function of $\sqrt{s}$ in Fig.\ref{R:tthh}. It shows the cross section in the TeV region is smaller than the SM prediction and can be changed to about $-34\%$, and the deviation decreases with the increasing  c.m. energy. These characteristics are similar with the $Z$ strahlung process.

In short, the double-$h$ productions can be useful to test the SM and further probe new physics beyond the SM via the $Z$ strahlung and $W$ boson fusion processes, but they show little sensitivity to $M_H$. The $t\bar{t}hh$ production is useful to extract top-Yukawa couplings and to check the self-consistency of Higgs scenario.

\section{Conclusions }

The EGH model is a perturbative extension of the SM featuring EGH boson and dark matter particle, which is theoretically well-motivated and phenomenologically viable. In this model, the EGH boson is taken as the observed Higgs boson, which almost has the same couplings with the SM fermions and the EW gauge bosons as those for the SM Higgs boson. Thus, this model can easily satisfy the experimental constraint from the Higgs signal data at the LHC.

The Higgs cubic self couplings play the most crucial role for the Higgs pair production.
The precise measurement of the SM-like Higgs cubic coupling can be achieved three modes in future high-energy $e^+e^-$ colliders: the $Z$ strahlung, $W$ boson fusion mechanism and associated production with top quarks.
In this paper, we have investigated pair production of the EGH boson at ILC and CLIC with the preferred values $\theta=0.018_{-0.003}^{+0.004}$ and $\alpha=1.57$ given by Ref.\cite{EGH2}. In the EGH model, for the $e^+e^-\rightarrow Zhh$, $e^+e^-\rightarrow \nu\bar{\nu}hh$, and $e^+e^-\rightarrow t\bar{t}hh$ processes, the cross sections can be changed about $-27\%$, $163\%$ and $-34\%$ at the c.m. energy of 500 GeV, 1.2 TeV and 1 TeV, respectively, with respect to those in the SM. According to the expected measurement precisions, such correction effects might be observed by the energy scan at the ILC and CLIC. We also find that since the trilinear self-coupling of the EGH boson is suppressed with respect to the SM one, the cross section of $h$ pair production is less sensitive to the mass parameter $M_H$ for all available modes.

We have compared the $\sqrt{s}$ dependence of the double SM-like Higgs boson production cross section in the EGH model with the predictions in other new physics models. In the two Higgs doublet model, the double Higgs boson production cross section via the $Z$ strahlung process can be enhanced by large enhancement of the triple Higgs boson coupling,  while the $e^+e^-\rightarrow \nu\bar{\nu}hh$ process can be suppressed \cite{eehh0}. In the EGH model, the deviations from the SM is not only on the triple Higgs boson coupling but also on the gauge couplings. The suppressed scale factors $K_{h[H]}^F$, $K_{h[H]}^V$, $\mu_h$ lead to the suppression of $Z$ strahlung process and the enhancement of $W$-fusion process. To compare with MCHMs, the cross section curves of the EGH model are presented in the same manner. We find that the ratios $R$ for the EGH model have more significant differences from the SM. In other words, the EGH model features larger deviations from the SM than the MCHM model, making the two models distinguishable.

In summary, at the ILC and the CLIC, we can expect that a large enhancement of the cross section of the $e^+e^-\rightarrow \nu\bar{\nu}hh$ process might be helpful to test the EGH model. Although the correction effects of the $e^+e^-\rightarrow Zhh$ and $e^+e^-\rightarrow t\bar{t}hh$ processes are relatively small, these processes are useful to check the self-consistency of Higgs scenario for EGH model. Furthermore, the measurements of the double EGH boson production processes at various $\sqrt{s}$ values would be helpful to discriminate EGH model from other new physics scenarios.

\section*{Acknowledgement}

\noindent
This work was supported in part by the National Natural Science Foundation of China under Grant No. 11275088 and Grants No. 11545012, the Natural Science Foundation of the Liaoning Scientific Committee (No.2014020151).

\end{document}